\documentstyle[preprint,tighten,aps,epsf,floats]{revtex}
\begin{document}

\title{First experimental results on creation and decay of
 $\eta$-mesic nuclei}

\author{G.A.\ Sokol%
\thanks{ Talk given at the IX Intern. Seminar ``Electromagnetic Interactions
of Nuclei at Low and Medium Energies", Sep 20-22, 2000, Moscow.
E-mail: gsokol@x4u.lebedev.ru},
T.A.\ Aibergenov, A.V.\ Koltsov, A.V.\ Kravtsov, \\
Yu.I.\ Krutov, A.I.\ L'vov, L.N.\ Pavlyuchenko, \\
V.P.\ Pavlyuchenko, S.S.\ Sidorin}

\address{P.N. Lebedev Physical Institute,
  Leninsky Prospect 53, Moscow 117924, Russia}

\maketitle

\begin{abstract}
First experimental results on photoproduction of $\eta$-mesic nuclei are
analyzed.  In an experiment performed at the 1 GeV electron synchrotron of
the Lebedev Physical Institute, correlated $\pi^+n$ pairs arising from the
reaction
$$
  \gamma + {}^{12}{\rm C} \to N + {}_\eta(A-1)
      \to N + \pi^+ + n + (A-2)
$$
and flying transversely to the photon beam have been observed.  When the
photon energy exceeds the $\eta$-meson production threshold, a distribution
of the $\pi^+n$ pairs over their total energy is found to have a peak in
the subthreshold region of the internal-conversion process $\eta p \to
\pi^+ n$ which signals about formation of $\eta$-mesic nuclei.
\end{abstract}

\bigskip

The idea that a bound state of the $\eta$-meson and a nucleus (the
so-called $\eta$-mesic nucleus) can exist in Nature was put forward long
ago by Peng \cite{pen85} who relied on the first estimates of the $\eta N$
scattering length $a_{\eta N}$ obtained by Bhalerao and Liu \cite{bha85}.
Owing to ${\rm Re\,} a_{\eta N} > 0$, an average attractive potential
exists between slow $\eta$ and nucleons. This can result in binding $\eta
A$ systems, provided the life time of $\eta$ in nuclei is long enough
\cite{liu86}.

Two attempts to discover $\eta$-nuclei were performed soon after
the first theoretical suggestions.  They were based on using
$\pi^+$-meson beams at BNL \cite{chr88} and LAMPF \cite{lei88} and both
failed, thus excluding properties of ${}_\eta A$ assumed in the first
works and challenged later \cite{chi88,kul98}.  A new interest in
studying the hypothetical $\eta$-nuclei arose out of an indirect
evidence for a formation of a quasi-bound $\eta$-${^3\rm He}$ state in
the reaction $pd \to \eta\, {^3\rm He}$ which would naturally explain
\cite{wil93,kon94} an experimentally observed near-threshold
enhancement in the total cross section of the reaction \cite{may96}.
A similar enhancement was found in the reaction $dd \to \eta \,{^4\rm He}$
as well \cite{wil97}.  Based on modern determinations of the $T$-matrix
of $\eta N$ scattering \cite{bat95,gre97}, theoretical calculations of the
$\eta$-nucleus scattering length $a_{\eta A}$ were fulfilled \cite{rak96}
which suggested that $\eta$-nuclei ${}_\eta A$ indeed exist
for all $A \ge 3$.

It should be kept in mind, however, that the above experimental
evidences from the reactions with $\eta$ in the final state do not
determine the {\em sign} of $a_{\eta A}$ \cite{may96} and thus cannot
unambiguously prove that $\eta$-nuclei really exist as bound rather
than virtual states.  Therefore, a crucial experiment would be in
an observation of bound $\eta$'s,
and the present work is aimed at doing that.

Specifically, in the present work a search for $\eta$-nuclei is
performed in the photo-mesonic reaction
\begin{equation}
    \gamma + {}^{12}\mbox{C} \to N + {}_\eta(A-1)
          \to N + \pi^+ n + (A-2),
\end{equation}
in which decay products of the $\eta$-nuclei are detected, viz.\
correlated pions and nucleons emitted in opposite directions
transversely to the beam. The underlying idea is \cite{sok91} that
such $\pi N$ pairs cannot be produced in quasi-free photoproduction
at energies as high as $E_\gamma \sim 700$ MeV, whereas they naturally
appear due to $\eta$'s stopped (captured) in the nucleus.
\begin{figure}[ht]
\epsfxsize=0.8\textwidth
\centerline{\epsfbox[23 656 428 787]{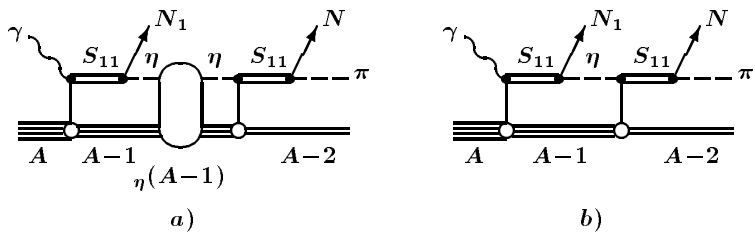}}
\caption{\sl
(a) Mechanism of formation and decay of an $\eta$-nucleus.
(b) Background production and decay of $\eta$'s in the nucleus.}
\end{figure}

The process of the $\eta$-nucleus formation in the reaction (1)
followed by the decay is shown schematically in Fig.~1a. There, both the
first stage of the reaction, i.e.\  production of $\eta$ by a photon,
and the second stage, i.e.\ annihilation of $\eta$ and creating a pion,
proceeds through single-nucleon interactions (either with a proton or a
neutron in the nucleus) mediated by the $S_{11}(1535)$ nucleon
resonance.  Formation of the bound state of the $\eta$ and the nucleus
becomes possible when the momentum of the produced $\eta$ is small
(typically less than 150 MeV/c, see below in Fig.~2).
This requirement suggests photon
energies $E_\gamma = 650{-}850$ MeV as most suitable for creating
$\eta$-nuclei.  Due to the Fermi motion, $\pi N$ pairs from
$\eta$-nuclei decays with characteristic opening angle
$\langle\theta_{\pi N}\rangle = 180^\circ$ have the width of $\simeq
25^\circ$. Their kinetic energies are $\langle E_\pi\rangle \simeq 300$
MeV and $\langle E_n\rangle \simeq 100$ MeV.
In the case when the momentum (or energy) of the
produced $\eta$ is high, the attraction between the $\eta$ and the
nucleus is not essential, and the $\eta$ propagates freely
(up to an absorption), see Fig.~1b.
In this case the final $\pi N$ pairs carry a high momentum too and
their kinematical characteristics, such as an opening angle, are
different from those of pairs produced through the stage of the
$\eta$-nucleus formation.

\begin{figure}[ht]
\epsfxsize=\textwidth
\epsfbox[68 568 558 728]{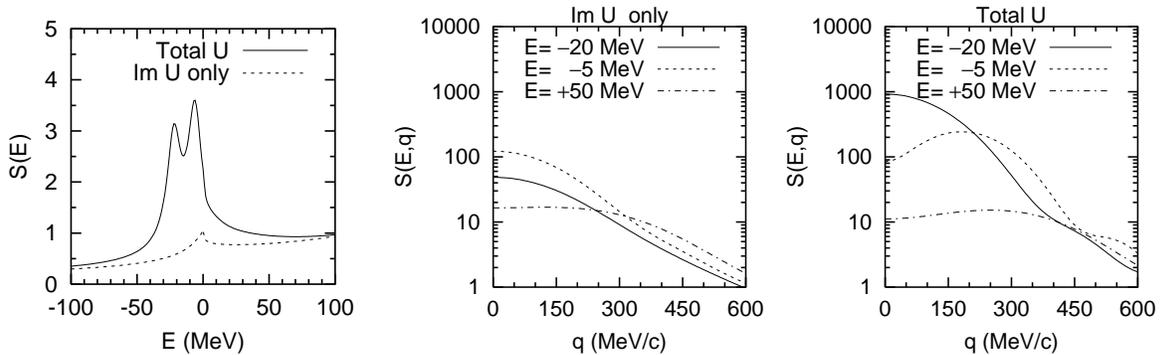}
\caption{\sl
Spectral functions $S(E)$ and $S(E,q)$ (in arbitrary units)
found with a rectangular-well optical potential
simulating the nucleus $^{12}$C.
For a comparison, shown also are results obtained with dropping out
the attractive (i.e.\ real) part of the $\eta A$ potential.}
\end{figure}

A systematic way to describe both the resonance (Fig.~1a) and
background (Fig.~1b) processes consists in using the Green function
$G({\bf r}_1,{\bf r}_2,E)$ which gives an amplitude of $\eta$ having an
energy $E$ to propagate between the creation and annihilation points in
the nuclear mean-field described by an optical energy-dependent
potential $U(r,E)$. In the vicinity of a bound level of a (complex)
energy $E_0$, the Green function has a pole $\sim 1/(E-E_0)$, and this
pole corresponds to the mechanism shown in Fig.~1a.  The
background process (Fig.~1b) corresponds to a non-pole part of
$G$. A convenient measure of the relative role of the background and
resonance processes is given by the spectral function
  $S(E) = \int \!\!\int \rho(r_1)
        \,\rho(r_2) \, |G({\bf r}_1,{\bf r}_2,E)|^2
        \,d{\bf r}_1 \, d{\bf r}_2$,
which characterizes a nuclear dependence of pion production through
the two-step transition $\gamma \to \eta \to \pi$ in the nucleus.
$S(E)$ depends on the binding potential $U$ and is proportional to
the number of $\eta N$ collisions which $\eta$ experiences
when travels through the nucleus of the density $\rho(r)$
between the creation and annihilation points.
An attractive potential $U$ makes the produced $\eta$ of a
near-resonance energy $E$ to pass several times through the nucleus
before it decays or escapes,
thus resulting in enhancing the number of collisions and in a resonance
increasing the production rate of the correlated $\pi N$ pairs.

A comparative role of the resonance and background contributions is
illustrated in Fig.~2 \cite{lvo98,sok98}, in which the spectral function
$S(E)$ is shown for the case of a rectangular-well optical potential $U$
simulating the $^{12}$C nuclear density and proportional to
the elementary $\eta N$-scattering amplitude by Green and Wycech
\cite{gre97}. The $\eta A$ attraction results in a prominent enhancement in
the number of collisions when $\eta$ has a negative energy in between
0 and $-30$ MeV.
A related spectral function
  $S(E,q) = \int \rho(r) \,
     |G_{r_2<R}({\bf r},{\bf q},E)|^2  \,d{\bf r}$,
which is given by Fourier components of the inner part of the Green function
(viz.\ a part having an overlap with nucleons in the nucleus),
describes a nuclear dependence of the energy-momentum
distribution $\partial^2 N / \partial E \partial{\bf q}$ of
the produced $\pi N$ pairs over their total energy and momentum
(which are $E + m_\eta + m_N$ and $q$, respectively,
up to the Fermi smearing).
As seen in Fig.~2, the $\eta$-nucleus attraction results
in a strong enhancement in the momentum density
at the resonance energies $E$ and low $q$.
This theoretical finding supports the starting point of the further
analysis that the correlated $\pi N$ pairs predominantly appear
from decays of bound $\eta$'s.

\begin{figure}[ht]
\leavevmode
\epsfxsize=0.30\textwidth\epsfbox[114 560 305 769]{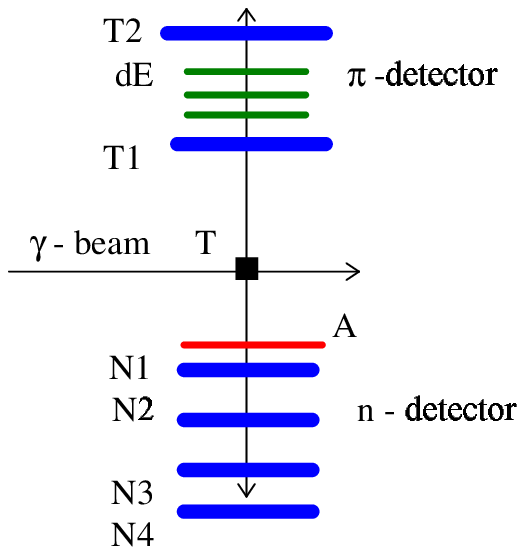}
\leavevmode
\epsfxsize=0.30\textwidth\epsfbox[000 000 184 220]{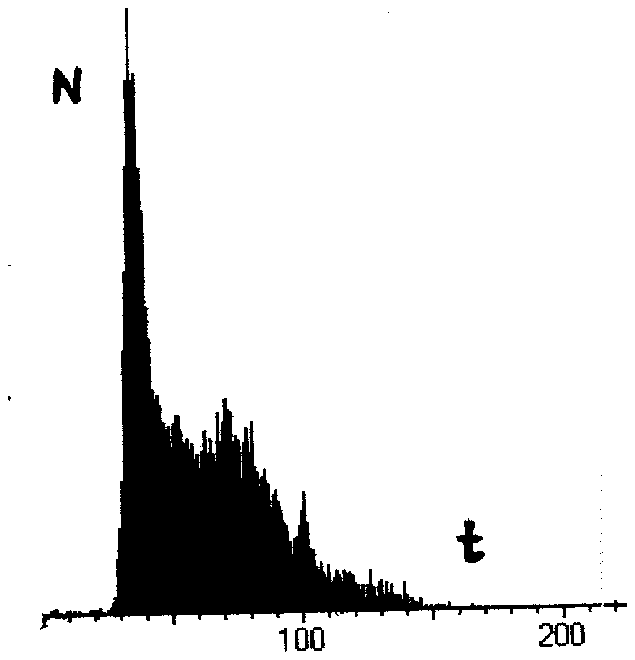}
\leavevmode
\epsfxsize=0.30\textwidth\epsfbox[000 000 200 210]{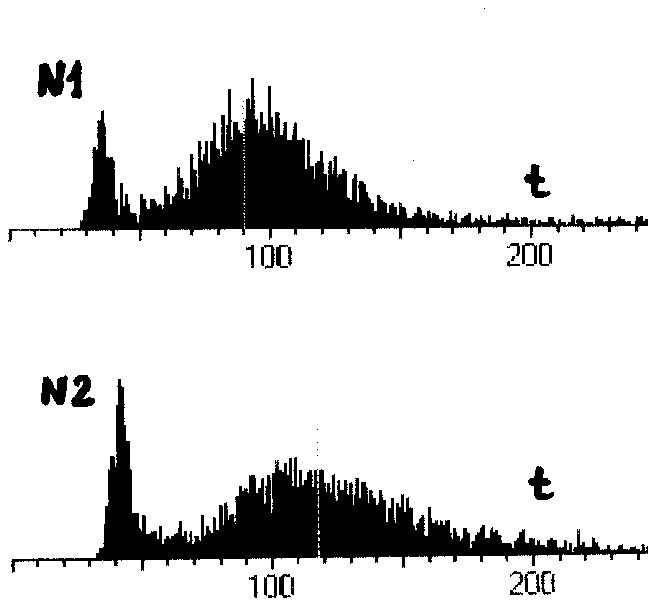}
\hfill
\bigskip
\caption{Layout of the experimental setup. Shown also
  time-of-flight spectra in the $\pi$ (left) and $n$ (right) spectrometers.}
\label{fig:setup}
\end{figure}

An experimental setup (Fig.~\ref{fig:setup})
consisted of a carbon target $\rm \, 4~cm \times 4~cm$ and two
time-of-flight scintillator spectrometers having a time resolution of
$\delta\tau \simeq 0.1$ ns. A plastic anticounter A of charged particles
(of the 90\% efficiency), placed in front of the neutron detectors,
and $dE/dx$ layers, placed between start and stop detectors in the pion
spectrometer, were used for a better identification of particles.

Strategy of measurements was as follows.
There were three runs in the present experiment with different
positions of the spectrometers: (a)  ``calibration", (b)
``background", and (c) ``effect $+$ background" runs.  In the
``calibration" run (a), both spectrometers were placed at
$\theta=50^\circ$ with respect to the photon beam, and the end-point
energy of the bremsstrahlung spectrum was $E_{\gamma \rm max}=650$ MeV.
In this run, mainly $\pi^+n$ pairs from quasi-free production of pions
from the carbon, $\gamma + {}^{12}\mbox{C} \to \pi^+ +  n + X$, were
detected.  In the ''background" run (b), the spectrometers were moved
at $\theta=90^\circ$ with respect to the photon beam, i.e.\ to the
position suitable for measuring the effect. However, the end-point
energy still was $E_{\gamma \rm max}=650$ MeV, i.e.\  well below the
$\eta$ photoproduction threshold off free nucleons (which is 707 MeV).
In the "effect$+$background" run (c), keeping the angle
$\theta=90^\circ$, the beam energy was set above the threshold:
$E_{\gamma \rm max}=850$ MeV.  The observed two-dimensional velocity
spectra of the detected pairs are shown in Fig.~4 for all three runs.

\begin{figure}[ht]
\epsfxsize=\textwidth
\epsfbox[23 440 470 800]{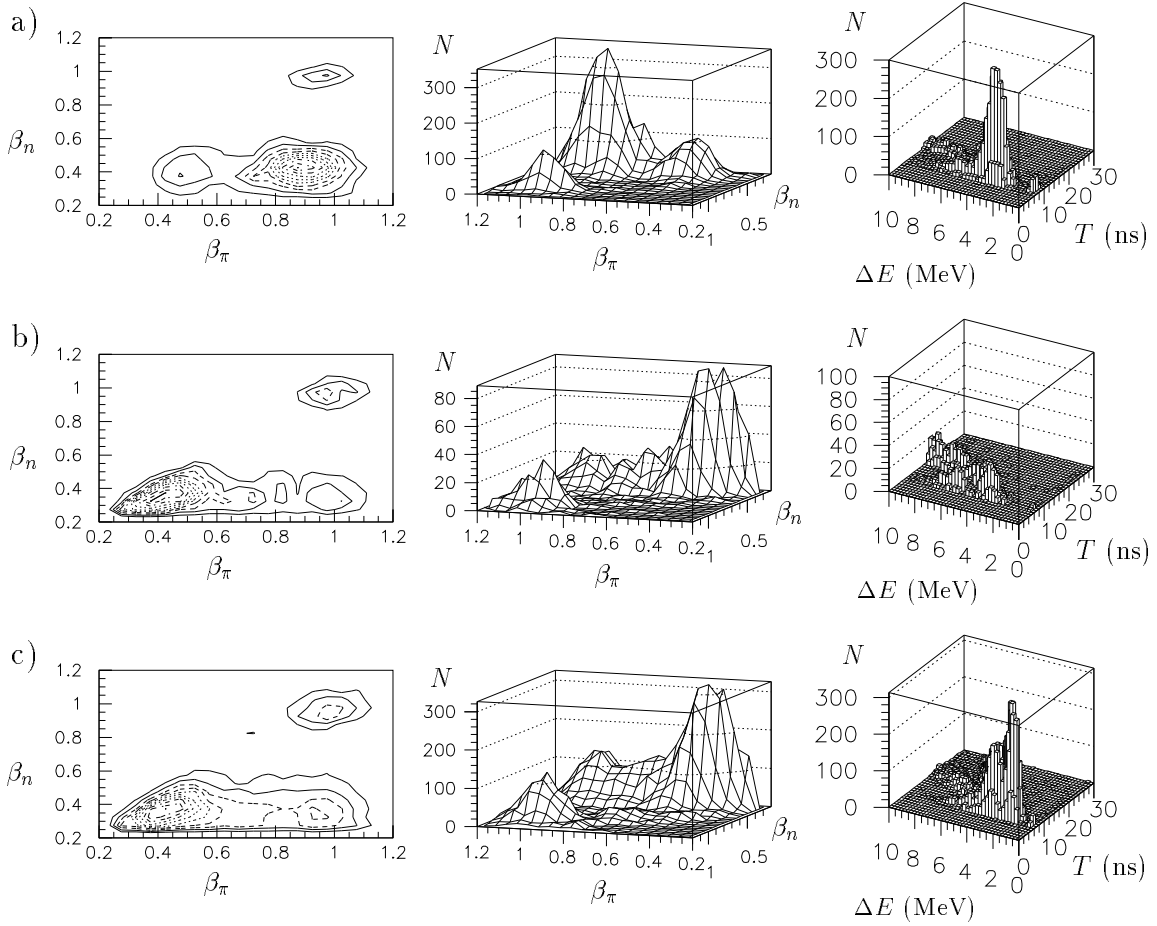}
\caption{\sl
Left and central panels:  Distributions (the number of events
$N$) over the pion and neutron velocities for the ``calibration" (a),
``background" (b), and ``effect$+$background" (c) runs.  Right panels:
Distributions over the time-of-flight $T$ and the energy losses $\Delta
E$ in the pion spectrometer for the same runs; only events with a slow
particle in the neutron spectrometer were selected for these plots.}
\end{figure}

In accordance with the velocities of particles in the pion and
neutron spectrometer, all events in each run can be assembled into
three groups:  fast-fast (FF), fast-slow (FS), and slow-slow (SS).  The
FF events with the extreme velocities close to the speed of the light
correspond to a background (mainly $e^+e^-$ pairs produced by $\pi^0$
from double-pion production).  The FS events mostly correspond to $\pi
N$ pairs.  In the ``calibration" run ($\theta=50^\circ$, $E_{\gamma \rm
max}=650$ MeV), the quasi-free production of the $\pi^+n$ pairs is seen
as a prominent peak in the two-dimensional distribution (Fig.~4a).  In
the ``background" run ($\theta=90^\circ$, $E_{\gamma \rm max}=650$
MeV), the largest peak (SS events in Fig.~4b) is caused by $\pi\pi$
pairs from double-pion photoproduction off the nucleus.  In the
``effect$+$background" run ($\theta=90^\circ$ and $E_{\gamma \rm
max}=850$ MeV) (Fig.~4c), apart from the SS events, a clear excess of
the FS events, as compared with the ``background" run, is seen.  This
FS signal is interpreted as a result of production and annihilation of
slow $\eta$'s in the nucleus giving the $\pi^+n$ pairs.

A further analysis of the events was done using an information from
three scintillation detectors which were positioned between the
start- and stop-layers of the time-of-flight pion spectrometer.
They measured the energy losses $\Delta E$ of particles passed through.
A selection of events with a minimal $\Delta E$ in two-dimensional
distributions over the time of flight $T$ and the energy losses $\Delta
E$ (Fig.~4) allows to discriminate events with a single pion from those
with the $e^+e^-$ pairs.

\newpage
The count rate of the $\pi^+n$ events was evaluated as
\begin{equation}
  N(\pi^+n; 850) =   N({\rm FS_{min}}; 850) - N({\rm FS_{min}}; 650)
    \times K(850/650),
\end{equation}
where $N({\rm FS_{min}};E_{\gamma \rm max})$ is the number of the
observed FS events with the minimal $\Delta E$ and with the specific
photon energy $E_{\gamma \rm max}$, and the coefficient K(850/650)
gives an increase of the FS-background due to double-pion
photoproduction when $E_{\gamma \rm max}$ grows from 650 MeV up to 850
MeV.  Assuming that the same coefficient describes an increase of the
SS count rate as well, it was found from the SS events at 650 and 850
MeV that $K=2.15$.  Such a procedure gives $N(\pi^+n; 850) = (61 \pm
7)$ events/hour.  Assuming an isotropic distribution of the $\pi^+n$
pairs, taking into account efficiencies of the pion and neutron
spectrometers (80\% and 30\%, respectively) and evaluating a
geometrical fraction $f$ of the correlated $\pi^+n$ pairs
simultaneously detected by the pion and neutron detectors of a finite
size (this fraction, $f=0.18$, was determined by a Monte Carlo
simulation of the width of the angular correlation between $\pi^+$ and
$n$ caused by the Fermi motion of nucleons and $\eta$ in the nucleus), we
obtain the following estimate of the total photoproduction cross
section of the correlated pairs from the carbon averaged over the
energy interval of 650--850 MeV:
\begin{equation}
\label{cstot}
  \langle \sigma(\pi^+n) \rangle =  (12.2 \pm 1.3) ~\mu\mbox{b}
\end{equation}
(a statistical error only).

Summarizing, we have observed a clear excess of the correlated $\pi^+n$
pairs with the opening angle close to $180^\circ$ arising when the
photon beam energy becomes higher than $\eta$-production threshold.
That is, we have observed production and decay of slow $\eta$'s inside
the nucleus.  As was discussed above in relation with Fig.~2,
these pairs are expected to be mostly related with a formation
and decay of $\eta$-nuclei in the intermediate state.
The obtained total cross section of pair production (\ref{cstot})
is close to theoretical predictions \cite{try95} for the total
cross section of $\eta$-nuclei formation in the photo-reaction,
what provides a further support for that expectation.

\begin{figure}[htb]
\centerline{\epsfxsize=0.8\textwidth
\epsfbox[37 260 548 582]{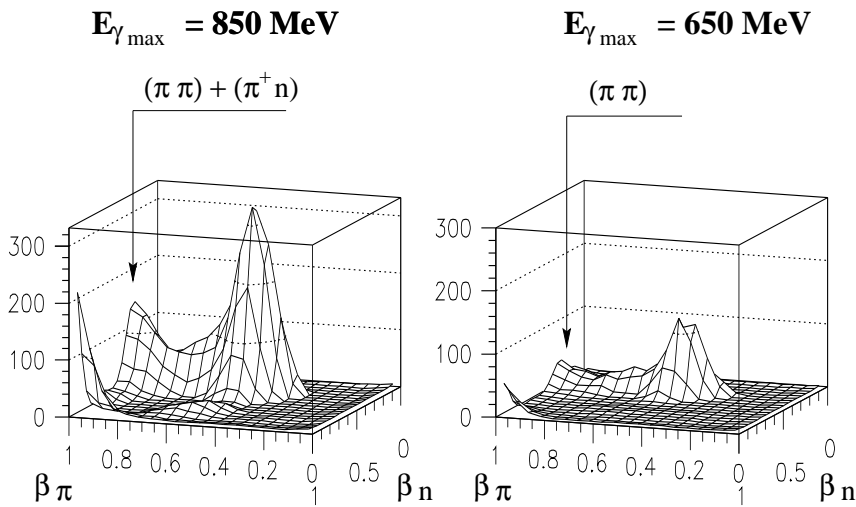}}
\caption{Corrected two-dimensional distributions over the velocities
    $\beta$ of the $\pi ^+ n$ events
    with the end-point energy of the bremsstrahlung spectrum
    $E_{\gamma \rm max} = 850$ and 650 MeV.}
\label{fig:betabeta-inverse}
\end{figure}

In a further analysis of the excess FS events, their 
energy characteristics have been studied.
In order to find kinetic energies of the neutron and pion, the velocities
$\beta_i = L_i/ct_i$ of both the particles must be determined.  They are
subject to fluctuations stemming from errors $\delta t_i$ and $\delta L_i$
in the time-of-flight $t_i$ and the flight base $L_i$.  Such fluctuations
are clearly seen in the case of the ultra-relativistic FF events which have
experimentally observed velocities close but not equal to 1 (see Fig.4
).  Therefore, an experimental $\beta$-resolution of
the setup can be directly inferred from the FF events.  Then, using this
information and applying an inverse-problem statistical method described in
Ref.\ \cite{pav83}, one can unfold the experimental
spectrum, obtain a
smooth velocity distribution in the physical region $\beta_i < 1$
(Fig.~\ref{fig:betabeta-inverse}), and eventually find a distribution of
the particle's kinetic energies $E_i = M_i [ (1-\beta_i^2)^{-1/2} - 1 ]$.
Finding $E_i$, we introduced corrections related with average energy losses
of particles in absorbers and in the detector matter.  It is worth to say
that the number of the $\pi^+ n$ FS events visibly increases when the
photon beam energy becomes sufficient for producing $\eta$ mesons.

\begin{figure}[htb]
\unitlength=1mm
\vspace*{10mm}
\begin{picture}(100,55)(0,0)
\put(40,55){\small $E_{\gamma \rm max}=850$ MeV}
\put(90,55){\small $E_{\gamma \rm max}=650$ MeV}
\centerline{\epsfxsize=5.5cm\epsfbox[157 322 443 613]{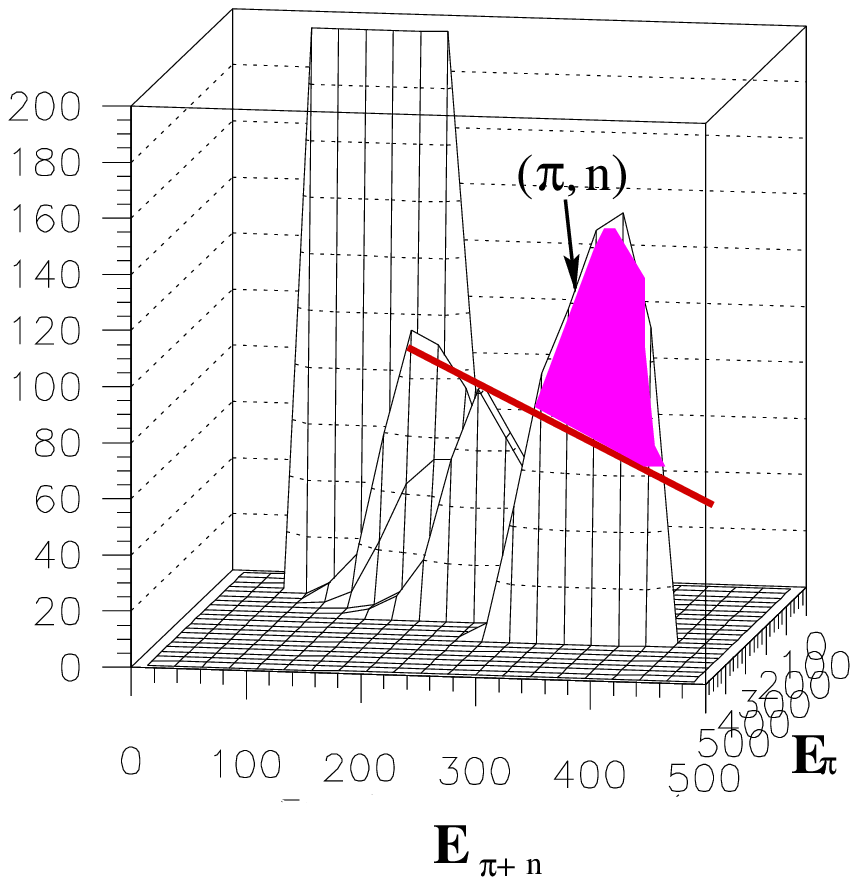}
\epsfxsize=4.5cm\epsfbox[16 325 304 670]{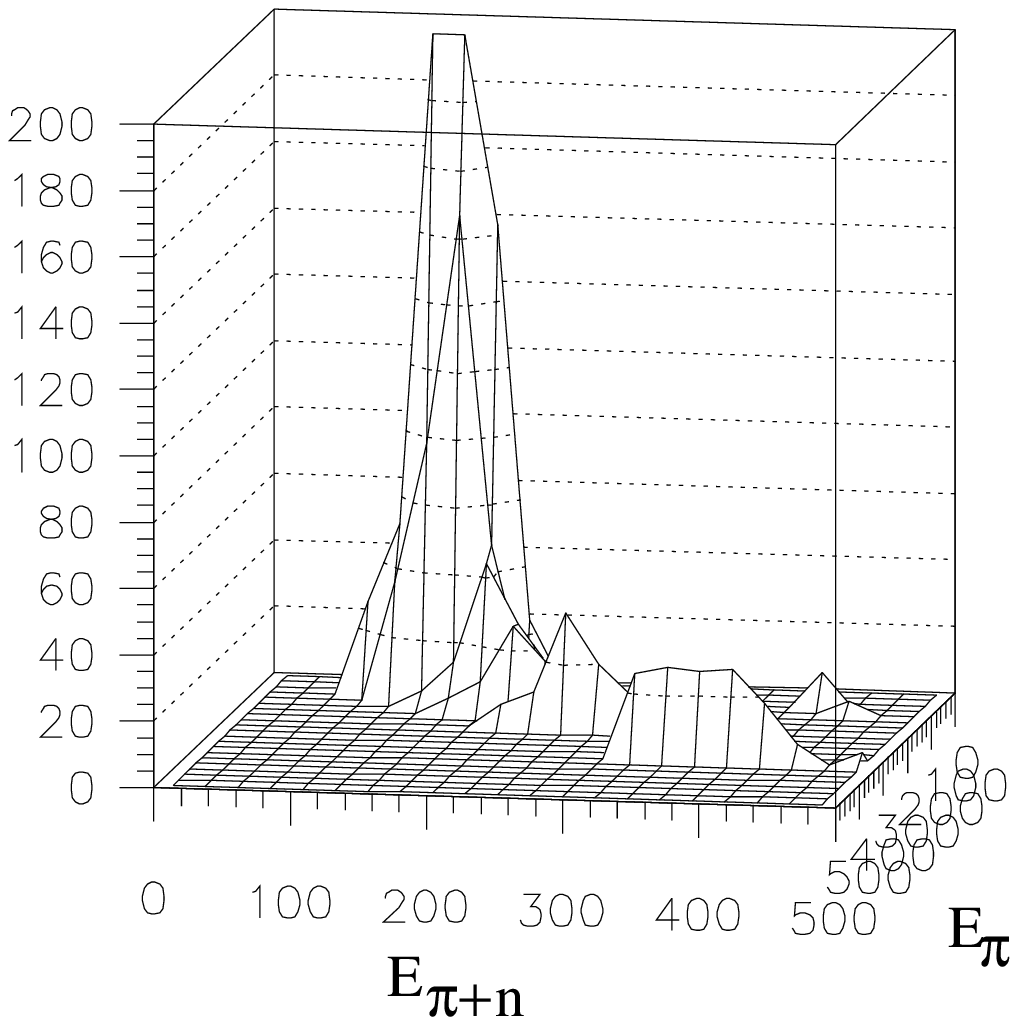}}
\end{picture}
\vspace*{5mm}
\caption{Distribution over the total kinetic energy of the $\pi^+n$ pairs
      for the ``effect$+$background" run (the left panel) and for the
      ``background" run (the right panel) obtained after
      unfolding the raw spectra.}
\label{fig:Etot-2dim}
\end{figure}

\begin{figure}[htb]
\vspace*{-10mm}
\centerline{\epsfxsize=0.45\textwidth
\epsfbox[43 18 340 298]{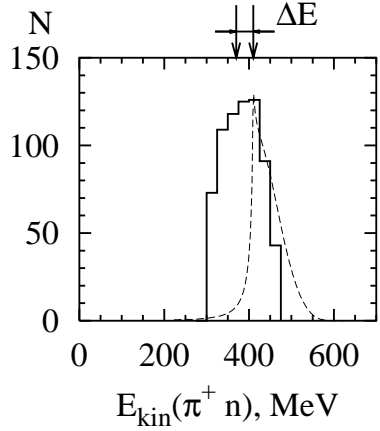}}
\caption{Distribution over the total kinetic energy of the $\pi^+n$ pairs
     after subtraction of the background. Arrows indicate threshold
     in the reaction $\eta N \to \pi N$, i.e.\ 408 MeV, and the weighted
     center of the histogram. For a comparison, a product of free-particle
     cross sections of $\gamma N \to \eta N$ and $\eta N \to \pi N$
     \protect\cite{gre97} is shown with the dashed line (in arbitrary units).}

\label{fig:Etot-1dim}
\end{figure}

Of the most interest is the distribution of the $\pi^+ n$ events over their
total energy $E_{\rm tot} = E_n + E_\pi$, because creation and decay of
$\eta$-mesic nuclei is expected to produce a relatively narrow peak in
$E_{\rm tot}$ of the width $\sim 50{-}70$ MeV (see, e.g.,
\cite{sok98,lvo98}). Such a peak was indeed observed: see Fig.\
\ref{fig:Etot-2dim}, in which an excess of the FS events appears when the
photon energy exceeds the $\eta$-production threshold. Subtracting a smooth
background, we have found a 1-dimensional energy distribution of the $\pi^+
n$ events presumably coming from (bound) $\eta$ decaying in the nucleus,
see Fig.\ \ref{fig:Etot-1dim}.

  The experimental width of this distribution
is about 100 MeV, including the apparatus resolution. Its center lies by
$\Delta E = 40$ MeV below the energy excess $m_\eta-m_\pi = 408$ MeV in the
reaction $\eta N \to \pi N$, and it is well below the position of the
$S_{11}(1535)$ resonance too.  Up to effects of binding of protons
annihilated in the decay subprocess $\eta p \to \pi^+ n$, the value $\Delta
E$ characterizes the binding energy of $\eta$ in the nucleus. The width of
that peak is determined both by the width of the $\eta$-bound state and by
the Fermi motion.\\

\begin{figure}[bht]
\unitlength=1mm
\begin{picture}(100,50)(0,0)
\put(030,30){\small $N$}
\put(115,01){\small $p_\perp$, MeV}
\put(042,45){\small $E_{\gamma \rm max}=850$ MeV}
\put(080,45){\small $E_{\gamma \rm max}=650$ MeV}
\centerline{\epsfxsize=0.5\textwidth
\epsfbox[17 390 542 690]{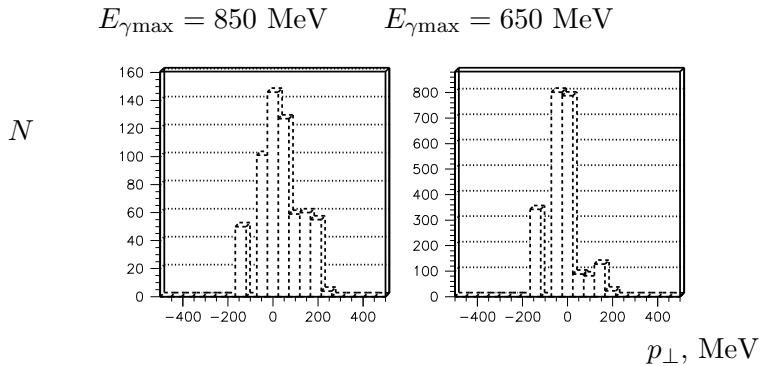}}
\end{picture}
\bigskip
\caption{Distribution over the total transverse momentum $p_\perp$ of $\pi^+n$
    pairs for the ``effect$+$background" run (the left panel)
    and for the ``calibration" run (the right panel).}
\label{fig:momentum}
\end{figure}

Whereas the fixed opening angle $\theta_{\pi n}=180^\circ$ chosen in the
kinematics with $\theta_n=\theta_\pi = 90^\circ$ selects $\pi^+n$ pairs
carrying a low total momentum in the direction of the photon beam, an
independent check of the transverse momentum $p_\perp = p_\pi - p_n$ is
meaningful.  The corresponding distribution is shown in Fig.\
\ref{fig:momentum}.  On the top of a background, there is a narrower peak
in $p_\perp$ having a width compatible with the Fermi momentum of nucleons
in the nucleus.

In conclusion, an excess of correlated $\pi^+ n$ pairs with the opening
angle $\langle \theta_{\pi N} \rangle = 180^\circ$ has been experimentally
observed when the energy of photons exceeded the $\eta$-production
threshold. A distribution of the pairs over their total kinetic energy was
found to have a peak lying below threshold of the elementary process $\pi
N\to \eta N$.  A narrow peak is also found in the pair's distribution over
their total transverse momentum. All that suggests that these $\pi^+ n$
pairs arise from creation and decay of captured bound $\eta$ in the
nucleus, i.e., they arise through the stage of formation of an $\eta$-mesic
nucleus.\\

\acknowledgments
The present research was supported by the Russian Foundation for
Basic Research, grants 96-02-17103, 99-02-18224.

\end{document}